\documentclass[iop]{emulateapj}

\usepackage{natbib,graphicx,amsmath}

\shorttitle{SECOND ORDER SOLUTIONS OF COSMOLOGICAL PERTURBATIONS}
\shortauthors{HWANG ET AL.}

\newcommand{\bea}{\begin{eqnarray}}
\newcommand{\eea}{\end{eqnarray}}

\begin{document}

\title{Second order solutions of cosmological perturbation in the matter dominated era}
\author{Jai-chan Hwang${}^{1,2}$, Hyerim Noh${}^{3}$ and Jinn-Ouk Gong${}^{4}$}
\address{${}^{1}$Department of Astronomy and Atmospheric Sciences,
                Kyungpook National University, Daegu 702-701, Republic of Korea \\
         ${}^{2}$Korea Institute for Advanced Study, Seoul 130-722, Republic of Korea
                \\
         ${}^{3}$Korea Astronomy and Space Science Institute, Daejeon 305-348,
                Republic of Korea \\
         ${}^{4}$Theory Division, CERN, CH-1211 Gen\`eve 23, Switzerland}

\begin{abstract}
We present the growing mode solutions of cosmological perturbations to the second order in the matter dominated era. We also present several gauge-invariant combinations of perturbation variables to the second order in most general fluid context. Based on the solutions we study the Newtonian correspondence of relativistic perturbations to the second order. In addition to the previously known exact relativistic/Newtonian correspondence of density and velocity perturbations to the second order in the comoving gauge, here we show that in the sub-horizon limit we have the correspondences for density, velocity and potential perturbations in the zero-shear gauge and in the uniform-expansion gauge to the second order. Density perturbation in the uniform curvature gauge also shows the correspondence to the second order in the sub-horizon scale. We also identify the relativistic gravitational potential which shows exact correspondence to the Newtonian one to the second order.
\end{abstract}


\section{Introduction}

Nonlinear evolution of perturbations is important to understand the observed large-scale structure in the universe.
Most of the study were made in Newtonian context (Peebles 1980; Vishniac 1983; Fry 1984; Goroff et al. 1986; Makino et al. 1992; Fry 1994; Bernardeau et al. 2002), but there have been growing interests in the corresponding studies in Einstein's gravity context (Tomita 1967, 1971, 1972; Kasai 1992, 1993; Bruni et al. 1997; Langlois \& Vernizzi 2005; Malik \& Matravers 2008; Malik \& Wands 2009; Langlois \& Vernizzi 2010; Nakamura 2010; Christopherson 2011).

Our aim in this work is to present a complete set of growing mode solutions of zero-pressure scalar-type perturbations to the second order. The solutions will be presented in a flat background with vanishing cosmological constant $K = 0 = \Lambda$. Using the solutions we can clarify the relativistic/Newtonian correspondences of the density, velocity and potential perturbations available in various gauges to the second order perturbations. Previously we have shown exact relativistic/Newtonian correspondence of the density and velocity perturbation variables to the second order in perturbation (Noh \& Hwang 2004, NH2004 hereafter), which is possible only in the comoving gauge condition.
In addition, the gauge transformation property of the metric and energy-momentum variables will be presented in the context of a general fluid with $K$ and $\Lambda$.

Section \ref{sec:perturbations} is a summary of our convention and basic equations. In Section \ref{sec:GT} we present the gauge transformation properties and some useful gauge-invariant combinations in general context. In Section \ref{sec:solutions} we show the growing mode solutions in various fundamental gauge conditions to the second order. In Section \ref{sec:correspondences} we discuss several distinguished relativistic/Newtonian correspondences. Section {\ref{sec:discussion} is a brief discussion.

The reader who is interested only in our conclusion about the relativistic/Newtonian correspondences can go directly to Section \ref{sec:correspondences}.

\section{Second-order perturbations}
                                     \label{sec:perturbations}

We consider scalar-type perturbations to the second order in Robertson-Walker background.
Our metric convention is (Bardeen 1988) \bea
   & & ds^2 = - a^2 \left( 1 + 2 \alpha \right) d \eta^2
       - 2 a^2 \beta_{,\alpha} d \eta d x^\alpha
   \nonumber \\
   & & \qquad
       + a^2 \left[ \left( 1 + 2 \varphi \right) g^{(3)}_{\alpha\beta}
       + 2 \gamma_{,\alpha|\beta} \right] d x^\alpha d x^\beta,
   \label{metric}
\eea where $a$ is the cosmic scale factor, and a vertical bar indicates a covariant derivative based on $g^{(3)}_{\alpha\beta}$. For the normalized fluid four-velocity we have $u_\alpha \equiv - a v_{,\alpha}$.
To the second order, the basic perturbation equations of general fluid\footnote{For the energy-momentum tensor to the second order, see Equations (54), (84) and (85) in Hwang \& Noh (2007).} are presented in Equations (95)-(103) of Hwang \& Noh (2007).  In the case of a zero-pressure ideal fluid the equations become
\bea
   & & \kappa - 3 H \alpha + 3 \dot \varphi + {\Delta \over a^2} \chi
       = n_0,
   \label{eq1} \\
   & & 4 \pi G \delta \mu + H \kappa + {\Delta + 3K \over a^2} \varphi
       = {1 \over 4} n_1,
   \label{eq2} \\
   & & \kappa
       + {\Delta + 3K \over a^2} \chi - 12 \pi G (\mu + p) a v
       = n_2,
   \label{eq3} \\
   & & \dot \kappa + 2 H \kappa
       - 4 \pi G \left( \delta \mu + 3 \delta p \right)
       + \left( 3 \dot H + {\Delta \over a^2} \right) \alpha
       = n_3,
   \label{eq4} \\
   & & \dot \chi + H \chi - \varphi - \alpha - 8 \pi G \Pi
       = n_4,
   \label{eq5} \\
   & & \delta \dot \mu + 3 H \left( \delta \mu + \delta p \right)
       - \left( \mu + p \right) \left( \kappa - 3 H \alpha
       + {1 \over a} \Delta v \right)
   \nonumber \\
   & & \qquad
       = n_5,
   \label{eq6} \\
   & & {[a^4 (\mu + p) v]^{\displaystyle\cdot} \over a^4(\mu + p)}
       - {1 \over a} \alpha
       - {1 \over a (\mu + p)} \left( \delta p
       + {2 \over 3} {\Delta + 3K \over a^2} \Pi \right)
   \nonumber \\
   & & \qquad
       = n_6,
   \label{eq7}
\eea where $n_i$ are quadratic combinations of linear order perturbations, and are presented in Equation (67)-(73) of Hwang \& Noh (2007); $\kappa$ is the perturbed part of the trace of extrinsic curvature,  $\chi \equiv a \beta + a^2 \dot \gamma$ is the shear of the normal frame vector, $\mu$ is the energy density, $p$ is the pressure, and $\Pi$ is the anisotropic stress; an overdot denotes a time derivative based on $t$ with $dt \equiv a d \eta$, and $H \equiv \dot a/a$. To the linear order, this set of equations is arranged in Bardeen (1988). The merits of this arrangement are (i) all perturbation variables are spatially gauge-invariant (this is true to the linear order, and can be extended to nonlinear order as well), and (ii) the temporal gauge (hypersurface or slicing condition) has not been taken, thus written in a sort of gauge-ready form. Setting any one of the perturbation variables equal to zero corresponds to a certain fundamental gauge condition: see below.

\section{Gauge transformation}
                                     \label{sec:GT}

In this section we consider a {\it general fluid} in the presence of $K$ and $\Lambda$ in the background.

\subsection{To the linear order}

We consider gauge transformation properties under $\widehat x^a = x^a + \widetilde \xi^a (x^e)$ with $\widetilde \xi^0 = \xi^0$, $\widetilde \xi^\alpha = \xi^\alpha$, and $\xi_\alpha \equiv \xi_{,\alpha}/a$; tildes indicate covariant quantities, whereas $\xi^\alpha$ is based on $g^{(3)}_{\alpha\beta}$.

To the linear order we have \bea
   & & \widehat \delta = \delta - {\mu^\prime \over \mu} \xi^0, \quad
       \widehat v = v - \xi^0, \quad
       \widehat \alpha = \alpha - {1 \over a} \left( a \xi^0 \right)^\prime,
   \nonumber \\
   & &
       \widehat \beta = \beta - \xi^0 + \left( {1 \over a} \xi \right)^\prime, \quad
       \widehat \gamma = \gamma - {1 \over a} \xi, \quad
       \widehat \varphi = \varphi - a H \xi^0,
   \nonumber \\
   & &
       \widehat \chi = \chi - a \xi^0, \quad
       \widehat \kappa = \kappa + \left( 3 \dot H
       + {\Delta \over a^2} \right) a \xi^0,
   \label{GT}
\eea where a prime denotes time derivative based on $\eta$. To the linear order, the following combinations are gauge invariant \bea
   & & \delta_v \equiv \delta + 3 \left( 1 + w \right) a H v
       \equiv 3 \left( 1 + w \right) a H v_\delta,
   \nonumber \\
   & &
       \varphi_v \equiv \varphi - a H v \equiv - a H v_\varphi, \quad
       \varphi_\chi \equiv \varphi - H \chi \equiv - H \chi_\varphi,
   \nonumber \\
   & &
       v_\chi \equiv v - {1 \over a} \chi \equiv - {1 \over a} \chi_v, \quad {\rm etc.},
\eea where $w \equiv p/\mu$. The gauge-invariant combination, for example, $\delta_v$ is the same as $\delta$ in the $v \equiv 0$ hypersurface condition, thus $\delta_v = \delta |_{v \equiv 0}$. The temporal gauge condition, for example, $v = 0$ fixes the temporal gauge mode completely. Thus, any perturbation variable in that gauge, for example $\delta$, can be equivalently regarded as a temporally gauge invariant ones, i.e., $\delta |_v = \delta_v$. Similar complete gauge fixing is true for the following temporal gauge conditions: the comoving gauge ($v \equiv 0$), the zero-shear gauge ($\chi \equiv 0$), the uniform-curvature gauge ($\varphi \equiv 0$), the uniform-expansion gauge ($\kappa \equiv 0$), and the uniform-density gauge ($\delta \equiv 0$). The synchronous gauge ($\alpha \equiv 0$) is an exception, leaving a remnant temporal gauge mode $\xi^t ({\bf x})$ even after fixing the gauge condition.

Similar complete gauge fixing and equivalence to unique gauge-invariant combination are possible for higher order perturbations as long as we take the spatial gauge condition in Equation (\ref{spatial-gauge}): see Section VI of NH2004. Concrete construction of the gauge-invariant combinations to the second order will be presented below.

\subsection{To the second order}

The gauge transformation properties of the fluid variables are\footnote{To the second order, the main fluid variables used in NH2004 were based on the normal frame. The energy frame fluid variables are also presented in Equation (238) of NH2004 from which we have Equations (\ref{GT-mu})-(\ref{GT-Pi}).} \bea
   & & \delta \widehat \mu
       = \delta \mu - \mu^\prime \xi^0
       - \delta \mu^\prime \xi^0
       + \mu^\prime \xi^0 \xi^{0 \prime}
       + {1 \over 2} \mu^{\prime\prime} \xi^0 \xi^0
   \nonumber \\
   & & \qquad
       - \left( \delta \mu
       - \mu^\prime \xi^0 \right)_{,\alpha} \xi^\alpha,
   \label{GT-mu} \\
   & & \delta \widehat p
       = \delta p
       - p^\prime \xi^0
       - \delta p^\prime \xi^0
       + p^\prime \xi^0 \xi^{0 \prime}
       + {1 \over 2} p^{\prime\prime} \xi^0 \xi^0
   \nonumber \\
   & & \qquad
       - \left( \delta p
       - p^\prime \xi^0 \right)_{,\alpha} \xi^\alpha,
   \label{GT-p} \\
   & & \widehat v
       = v - \xi^0
       + \xi^0 \left( \xi^{0 \prime} + {1 \over 2} {a^\prime \over a} \xi^0 \right)
       - \left( v - \xi^0 \right)_{,\alpha} \xi^\alpha
   \nonumber \\
   & & \qquad
       - \Delta^{-1} \nabla^\alpha \left[
       \alpha \xi^0_{\;,\alpha}
       + \left( v^\prime + {a^\prime \over a} v \right)_{,\alpha} \xi^0 \right],
   \label{GT-v} \\
   & & \widehat \Pi_{\alpha\beta}
       = \Pi_{\alpha\beta}
       - \left( \Pi_{\alpha\beta}^\prime
       + 2 {a^\prime \over a} \Pi_{\alpha\beta}  \right) \xi^0
       - \Pi_{\alpha\beta,\gamma} \xi^\gamma
   \nonumber \\
   & & \qquad
       - 2 \Pi_{\gamma(\alpha} \xi^\gamma_{\;\;,\beta)}.
   \label{GT-Pi}
\eea Gauge transformation properties of metric perturbations are presented in Equation (278) of NH2004. 

Now we address the issue of spatial gauge condition.
The gauge conditions include the hypersurface (temporal gauge) condition and congruence (spatial gauge) condition. To the linear order, the basic equations are already written in a spatially gauge-invariant form without directly involving $\beta$ or $\gamma$ separately (Bardeen 1988). All perturbation variables in Equations (\ref{eq1})-(\ref{eq7}) are spatially gauge-invariant or equivalently the same as the variables in the $\gamma = 0$ spatial gauge; this choice is unique because taking $\beta = 0$ will leave a remnant gauge mode (Bardeen 1988). Similarly, now to the second order we will take \bea
   & & \gamma \equiv 0,
   \label{spatial-gauge}
\eea as the spatial gauge (threading or congruence) condition. The choice is unique in the following sense. As $\gamma = 0$ completely fixes the spatial gauge mode, the remaining equations after fixing this condition can be equivalently regarded as spatially gauge-invariant ones even to the nonlinear order (Bardeen 1988; NH2004). Also our gauge related properties do not depend on our ignoring the vector and tensor type perturbations which are coupled to the nonlinear order (NH2004). Under this spatial gauge condition we have $\xi^\alpha = 0$ even to the second order.

From Equations (\ref{GT-mu})-(\ref{GT-Pi}) above, and Equation (278) in NH2004, we find the gauge transformation properties of the fluid and metric variables as\footnote{For $\kappa$ there is a typo in the sign of last term of Equation (279) in NH2004.} \begin{widetext} \bea
   \delta \widehat \mu
   &=& \delta \mu - \mu^\prime \xi^0
       - \delta \mu^\prime \xi^0
       + \mu^\prime \xi^0 \xi^{0 \prime}
       + {1 \over 2} \mu^{\prime\prime} \xi^0 \xi^0,
   \label{GT-mu-2} \\
   \delta \widehat p
   &=& \delta p
       - p^\prime \xi^0
       - \delta p^\prime \xi^0
       + p^\prime \xi^0 \xi^{0 \prime}
       + {1 \over 2} p^{\prime\prime} \xi^0 \xi^0,
   \label{GT-p-2} \\
   \widehat v
   &=& v - \xi^0
       + \xi^0 \left( \xi^{0 \prime} + {1 \over 2} {a^\prime \over a} \xi^0 \right)
       - \Delta^{-1} \nabla^\alpha \left[
       \alpha \xi^0_{\;,\alpha}
       + \left( v^\prime + {a^\prime \over a} v \right)_{,\alpha} \xi^0 \right],
   \label{GT-v-2} \\
   \widehat \Pi_{\alpha\beta}
   &=& \Pi_{\alpha\beta}
       - \left( \Pi_{\alpha\beta}^\prime
       + 2 {a^\prime \over a} \Pi_{\alpha\beta}  \right) \xi^0,
   \label{GT-Pi-2} \\
   \widehat \alpha
   &=& \alpha - {1 \over a} \left( a \xi^0 \right)^\prime
       - \alpha^\prime \xi^0
       - 2 \alpha {1 \over a} \left( a \xi^0 \right)^\prime
       + \xi^0 \xi^{0 \prime\prime}
       + {3 \over 2} \xi^{0\prime} \xi^{0\prime}
       + 3 {a^\prime \over a} \xi^0 \xi^{0\prime}
       + {1 \over 2} \left(
       {a^{\prime\prime} \over a}
       + {a^{\prime 2} \over a^2} \right) \xi^0 \xi^0,
   \\
   \widehat \varphi
   &=& \varphi
       - {a^\prime \over a} \xi^0
       + \left[ - \varphi^\prime
       - 2 {a^\prime \over a} \varphi
       + {a^\prime \over a} \xi^{0\prime}
       + {1 \over 2} \left( {a^{\prime\prime} \over a}
       + {a^{\prime 2} \over a^2} \right) \xi^0 \right] \xi^0
   \nonumber \\
   & &
       + {1 \over 2} \left(
       {1 \over a} \chi^{,\alpha} \xi^0_{\;,\alpha}
       - {1 \over 2} \xi^{0,\alpha} \xi^0_{\;,\alpha} \right)
       - {1 \over 2} \Delta^{-1} \nabla^\alpha \nabla^\beta \left(
       {1 \over a} \chi_{,\alpha} \xi^0_{\;,\beta}
       - {1 \over 2} \xi^0_{\;,\alpha} \xi^0_{\;,\beta} \right),
   \\
   \widehat \chi
   &=& \chi
       - a \xi^0
       + \xi^0 \left( a \xi^0 \right)^\prime
       + \Delta^{-1} \nabla^\alpha \left[
       - {1 \over a} \left( a \chi_{,\alpha} \xi^0 \right)^\prime
       + \left( - 2 \alpha + \xi^{0 \prime} \right)
       a \xi^0_{\;,\alpha} \right]
   \nonumber \\
   & &
       - {a \over 2} \Delta^{-1} \left[
       {1 \over a}
       \chi^{,\alpha} \xi^0_{\;,\alpha}
       - {1 \over 2} \xi^{0,\alpha} \xi^0_{\;,\alpha}
       - 3 \Delta^{-1} \nabla^\alpha \nabla^\beta \left(
       {1 \over a} \chi_{,\alpha} \xi^0_{\;,\beta}
       - {1 \over 2} \xi^0_{\;,\alpha} \xi^0_{\;,\beta} \right) \right]^\prime,
   \\
   \widehat \kappa
   &=& \kappa
       + \left( 3 \dot H + {\Delta \over a^2} \right) a \xi^0
       + \left[ - \kappa^\prime
       - \left( 3 \dot H + {\Delta \over a^2} \right) \left( a \xi^0 \right)^\prime
       + {3 \over 2} \left( H \dot H - \ddot H \right) a^2 \xi^0
       \right] \xi^0
   \nonumber \\
   & &
       + {1 \over a} \left( 2 \alpha + \varphi
       - 2 \xi^{0 \prime}
       - {3 \over 2} {a^\prime \over a} \xi^0 \right)^{,\alpha} \xi^0_{\;,\alpha}
       + \left( \alpha - 2 \varphi - \xi^{0 \prime}
       + 2 {a^\prime \over a} \xi^0 \right) {\Delta \over a} \xi^0.
\eea

\subsection{Gauge-invariant combinations}
                                        \label{sec:GI}

From the gauge transformation properties of the metric and fluid variables we can construct the following gauge-invariant combinations\footnote{For method, see Section VI.C.2 of NH2004. The combination $\varphi_\chi$ was presented in Equation (280) of NH2004; the other ones in NH2004 are based on the fluid quantities in the normal frame, thus here we are presenting again in the energy frame.}
\bea
   \varphi_v
   &\equiv&
       \varphi - a H v
       - a v \left[ \dot \varphi_v + 2 H \varphi_v
       + {1 \over 2} \left( \dot H + 2 H^2 \right) a v \right]
   \nonumber \\
   & &
       + {1 \over 2} \left( {1 \over a} \chi^{\;,\alpha}_v v_{,\alpha}
       + {1 \over 2} v^{,\alpha} v_{,\alpha} \right)
       - {1 \over 2} \Delta^{-1} \nabla^\alpha \nabla^\beta
       \left( {1\over a} \chi_{v,\alpha} v_{,\beta}
       + {1 \over 2} v_{,\alpha} v_{,\beta} \right)
       + a H \Delta^{-1} \nabla^\alpha \left( \alpha_v v_{,\alpha} \right),
   \\
   \varphi_\chi
   &\equiv&
       \varphi - H \chi
       - \left( \dot \varphi_\chi + 2 H \varphi_\chi \right) \chi
       - {1 \over 2} \left( \dot H + H^2 \right) \chi^2
       + {1 \over 4 a^2} \left[ \chi^{,\alpha} \chi_{,\alpha}
       - \Delta^{-1} \nabla^\alpha \nabla^\beta
       \left( \chi_{,\alpha} \chi_{,\beta} \right) \right]
   \nonumber \\
   & &
        + H \Delta^{-1} \nabla^\alpha \left[ 2 \alpha_\chi \chi_{,\alpha}
        + \left( \dot \chi - H \chi \right) \chi_{,\alpha} \right]
        + {1 \over 4} a^2 H \Delta^{-1} \left[
        {1 \over a^2} \chi^{,\alpha} \chi_{,\alpha}
        - 3 {1 \over a^2} \Delta^{-1} \nabla^\alpha \nabla^\beta
        \left( \chi_{,\alpha} \chi_{,\beta} \right) \right]^{\displaystyle\cdot},
   \label{GI-varphi-chi}
\eea
\bea
   \varphi_\delta
   &\equiv&
       \varphi
       + {\delta \over 3 ( 1 + w )}
       + \left[ {1 \over H} \dot \varphi_\delta
       + 2 \varphi_\delta
       + \left( - 1 - {1 \over 2} {\dot H \over H^2}
       + {1 \over 2} {\ddot \mu \over H \dot \mu} \right)
       {\delta \over 3 ( 1 + w )} \right]
       {\delta \over 3 (1 + w)}
   \nonumber \\
   & &
       + {1 \over a^2} {1 \over 6 (1 + w) H} \left\{
       \left( - \chi_\delta
       + {\delta \over 6 (1+w)H} \right)^{,\alpha} \delta_{,\alpha}
       - \Delta^{-1} \nabla^\alpha \nabla^\beta \left[
       \left( - \chi_{\delta} + {\delta \over 6 (1+w) H}
       \right)_{,\alpha} \delta_{,\beta}
       \right] \right\},
   \label{varphi_delta}
   \\
   \delta_\varphi
   &\equiv&
       \delta + 3 \left( 1 + w \right) \varphi
       + \left[ - {\delta \dot \mu_\varphi \over \mu}
       + \left( {\dot \mu \over \mu}
       + {1 \over 2} {\dot \mu \dot H \over \mu H^2}
       - {1 \over 2} {\ddot \mu \over \mu H} \right) \varphi
       \right] {\varphi \over H}
   \nonumber \\
   & &
        + {3 (1 + w) \over 2 a^2}
        \left\{ \left( \chi_\varphi + {\varphi \over 2H} \right)^{,\alpha}
        {\varphi_{,\alpha} \over H}
        - \Delta^{-1} \nabla^\alpha \nabla^\beta
        \left[ \left( \chi_\varphi + {\varphi \over 2H} \right)_{,\alpha} {\varphi_{,\beta} \over H} \right] \right\},
   \label{delta_varphi} \\
   \delta_v
   &\equiv&
       \delta - {\dot \mu \over \mu} a v
       - \left( {\delta \dot \mu_v \over \mu}
       + {1 \over 2} {\ddot \mu \over \mu} a v \right) a v
       + {\dot \mu \over \mu} \Delta^{-1} \nabla^\alpha
       \left( \alpha_v a v_{,\alpha} \right),
   \label{delta_v} \\
   \delta p_v
   &\equiv&
       \delta p - \dot p a v
       - \left( \delta \dot p_v
       + {1 \over 2} \ddot p a v \right) a v
       + \dot p \Delta^{-1} \nabla^\alpha
       \left( \alpha_v a v_{,\alpha} \right),
   \label{p_v} \\
   v_\chi
   &\equiv&
       v - {1 \over a} \chi
       + {1 \over a} \Delta^{-1} \nabla^\alpha \left[
       \alpha_\chi \chi_{,\alpha}
       - \chi \left( a v_\chi \right)^{\displaystyle\cdot}_{,\alpha}
       + \dot \chi \chi_{,\alpha} \right]
       + {a \over 4} \Delta^{-1} \left\{
       {1 \over a^2} \left[
       \chi^{,\alpha} \chi_{,\alpha}
       - 3 \Delta^{-1} \nabla^\alpha \nabla^\beta \left(
       \chi_{,\alpha} \chi_{,\beta} \right) \right] \right\}^{\displaystyle\cdot},
   \\
   \chi_v
   &\equiv&
       \chi - a v
       - {1 \over 2} a^2 H v^2
       - a^2 \Delta^{-1} \left( \dot v \Delta v
       + \dot v^{,\alpha} v_{,\alpha} \right)
       - a \Delta^{-1} \nabla^\alpha \left[
       \alpha_v v_{,\alpha}
       + {1 \over a} v \left( a \chi_v \right)^{\displaystyle\cdot}_{,\alpha}
       + \dot v \chi_{v,\alpha} \right]
   \nonumber \\
   & &
       - {a^2 \over 2} \Delta^{-1} \left[
       {1 \over a} \chi_v^{\;,\alpha} v_{,\alpha}
       + {1 \over 2} v^{,\alpha} v_{,\alpha}
       - 3 \Delta^{-1} \nabla^\alpha \nabla^\beta
       \left( {1 \over a} \chi_{v,\alpha} v_{,\beta}
       + {1 \over 2} v_{,\alpha} v_{,\beta} \right) \right]^{\displaystyle\cdot},
   \\
   \kappa_v
   &\equiv&
       \kappa + \left( 3 \dot H + {\Delta \over a^2} \right) a v
       - a \dot \kappa_v v
       + {3 \over 2} a^2 \ddot H v^2
       - 2 \left( \varphi_v + a H v \right) {\Delta \over a} v
       + {1 \over a} \left( \alpha_v + \varphi_v
       + {1 \over 2} a H v \right)^{,\alpha} v_{,\alpha}
   \nonumber \\
   & &
       - 3 a \dot H \Delta^{-1} \left(
       \alpha_v \Delta v
       + \alpha_v^{\;,\alpha} v_{,\alpha} \right).
   \label{kappa_v}
\eea We introduced the entropic perturbation $e$ as \bea
   & & e \equiv \delta p - {\dot p \over \dot \mu} \delta \mu,
\eea which is gauge-invariant only to the linear order. We can show that \bea
   e_v \equiv
       e - \dot e a v
       - \left( \ddot p - {\dot p \over \dot \mu} \ddot \mu \right)
       \left( {\delta \mu_v \over \dot \mu}
       + {1 \over 2} a v \right) a v.
   \label{e_v}
\eea Gauge-invariant combinations involving $\kappa$ are somewhat complicated. For example, we have \bea
   \left( 3 \dot H + {\Delta \over a^2} \right) \varphi_\kappa
   &\equiv&
       \left( 3 \dot H + {\Delta \over a^2} \right) \varphi + H \kappa
       + \left[ \left( 3 \dot H + {\Delta \over a^2} \right) \varphi_\kappa \right]^{\displaystyle\cdot}
       {\kappa \over 3 \dot H + {\Delta \over a^2}}
       - \left[ \left( 3 \ddot H - 6 H \dot H - 4 H {\Delta \over a^2} \right) \varphi_\kappa
       \right] {\kappa \over 3 \dot H + {\Delta \over a^2}}
   \nonumber \\
   & &
       - 3 \left( H^2 \dot H + {1 \over 2} \dot H^2
       - {1 \over 2} H \ddot H \right) \left( {\kappa \over 3 \dot H + {\Delta \over a^2}} \right)^2
       + {1 \over a^2} \left(
       2 \dot \varphi_\kappa
       + 3 H \varphi_\kappa
       - 2 H \alpha_\kappa
       \right)^{,\alpha}
       {\kappa_{,\alpha} \over 3 \dot H + {\Delta \over a^2}}
   \nonumber \\
   & &
       - {1 \over a^2} \left( \dot H + {3 \over 2} H^2 \right)
       {\kappa^{,\alpha} \over 3 \dot H + {\Delta \over a^2}}
       {\kappa_{,\alpha} \over 3 \dot H + {\Delta \over a^2}}
       + \left( \dot \varphi_\kappa + 4 H \varphi_\kappa - H \alpha_\kappa \right)
       {\Delta \over a^2} {\kappa \over 3 \dot H + {\Delta \over a^2}}
   \nonumber \\
   & &
       - \left( \dot H + 4 H^2 \right)
       {\kappa \over 3 \dot H + {\Delta \over a^2}}
       {\Delta \kappa \over 3 \dot H + {\Delta \over a^2}}
       + \left( 3 \dot H + {\Delta \over a^2} \right)
       {1 \over 2 a^2} \Bigg[
       - \chi_\kappa^{\;,\alpha}
       {\kappa_{,\alpha} \over 3 \dot H + {\Delta \over a^2}}
       + {1 \over 2}
       {\kappa^{,\alpha} \over 3 \dot H + {\Delta \over a^2}}
       {\kappa_{,\alpha} \over 3 \dot H + {\Delta \over a^2}}
   \nonumber \\
   & &
       - \Delta^{-1} \nabla^\alpha \nabla^\beta \left(
       - \chi_{\kappa,\alpha}
       {\kappa_{,\beta} \over 3 \dot H + {\Delta \over a^2}}
       + {1 \over 2}
       {\kappa_{,\alpha} \over 3 \dot H + {\Delta \over a^2}}
       {\kappa_{,\beta} \over 3 \dot H + {\Delta \over a^2}}
       \right) \Bigg],
   \label{varphi_kappa}
\eea \end{widetext} where $ \kappa / (3 \dot H + \Delta / a^2) =  [(3 \dot H + \Delta / a^2)^{-1} \kappa]$. These gauge-invariant combinations will be used to relate solutions in different gauges in the next section. .

\section{Growing mode solutions in matter dominated era}
                                     \label{sec:solutions}

Now we consider a matter dominated era with $K = 0 = \Lambda$. In this section we will present growing mode solutions of a zero-pressure fluid to the second order. Complete solutions to the linear order are presented in Hwang (1994).

\subsection{Comoving gauge}

We take the comoving gauge \bea
   & & v \equiv 0 \equiv \gamma.
\eea

To the second-order perturbation, from Equations (\ref{eq1}), (\ref{eq3}) and (\ref{eq7}) we have \bea
   & & \dot \varphi_v
       = {1 \over 3} \left( n_0 |_v - n_2 |_v \right)
       - {a H \over c^2} n_6 |_v.
   \label{dot-varphi-v}
\eea
To the linear order we have simply \bea
   & & \dot \varphi_v = 0.
\eea
Thus $\varphi_v = C({\bf x})$ with vanishing transient solution where $C({\bf x})$ is an integration constant of the growing solution.
To the linear order, the relatively growing solutions in expanding phase are \bea
   & & \varphi_v = C, \quad
       \alpha_v = 0, \quad
       \chi_v = {2 \over 5} {1 \over H} C, \quad
       \kappa_v = - {2 \over 5} {\Delta \over a^2 H} C,
   \nonumber \\
   & &
       \delta_v = - {2 \over 5} {\Delta \over a^2 H^2} C.
\eea
By plugging these linear solutions into $n_i$ of Equations (\ref{eq1})-(\ref{eq7}), we can obtain the full second order solutions: Equation (\ref{eq7}) gives $\alpha_v$; Equations (\ref{eq1}) and (\ref{eq3}) give $\varphi_v$; Equation (\ref{eq5}) gives $\chi_v$; Equation (\ref{eq3}) gives $\kappa_v$; and finally Equation (\ref{eq2}) gives $\delta_v$. Then, we have \begin{widetext} \bea
   \alpha_v
   &=& - {2 \over 25} {1 \over a^2 H^2} C^{,\alpha} C_{,\alpha},
   \nonumber \\
   \varphi_v
   &=& C - {1 \over 5} {1 \over a^2 H^2} \left[
       {1 \over 2} C^{,\alpha} C_{,\alpha}
       + \Delta^{-1} \nabla^\alpha \left( C_{,\alpha} \Delta C \right) \right],
   \nonumber \\
   \chi_v
   &=& {1 \over 5} {1 \over H} \left[ 2 C - C^2
       + \Delta^{-1} \left( C \Delta C \right)
       - 3 \Delta^{-1} \Delta^{-1} \nabla^\alpha \nabla^\beta
       \left( C C_{,\alpha\beta} \right) \right]
       - {4 \over 175} {1 \over a^2 H^3} \left[
       2 C^{,\alpha} C_{,\alpha}
       + 3 \Delta^{-1} \nabla^\alpha \left( C_{,\alpha} \Delta C \right) \right],
   \nonumber \\
   \kappa_v
   &=& {1 \over 5} {1 \over a^2 H} \left( - 2 \Delta C
       + 8 C \Delta C
       + 3 C^{,\alpha} C_{,\alpha} \right)
       + {4 \over 175} {1 \over a^4 H^3}
       \left[ 2 \Delta \left( C^{,\alpha} C_{,\alpha} \right)
       + 3 \nabla^\alpha \left(
       C_{,\alpha} \Delta C \right) \right],
   \nonumber \\
   \delta_v
   &=& {1 \over 5} {1 \over a^2 H^2} \left( - 2 \Delta C
       + 8 C \Delta C
       + 3 C^{,\alpha} C_{,\alpha} \right)
       + {4 \over 175} {1 \over a^4 H^4}
       \left[ \Delta \left( C^{,\alpha} C_{,\alpha} \right)
       + 5 \nabla^\alpha \left(
       C_{,\alpha} \Delta C \right) \right].
   \label{CG-sol}
\eea \end{widetext}
These are a complete set of solutions in the comoving gauge.
Using this set of solutions we can derive solutions in any of other gauge conditions.

\subsection{Zero-shear gauge}

The zero-shear gauge condition sets \bea
   & & \chi \equiv 0 \equiv \gamma.
\eea This gauge is often referred to as the conformal Newtonian or longitudinal gauge.
To the linear order we have \bea
   & & \varphi_\chi = {3 \over 5} C, \quad
       \alpha_\chi = - {3 \over 5} C, \quad
       \kappa_\chi = - {9 \over 5} H C,
   \nonumber \\
   & &
       v_\chi = - {2 \over 5} {1 \over a H} C, \quad
       \delta_\chi = {2 \over 5} \left( 3 - {\Delta \over a^2 H^2} \right) C.
\eea
Using these, we can calculate $n_i$ in the zero-shear gauge.
Evaluating Equation (\ref{GI-varphi-chi}) in the comoving gauge
we can derive $\varphi_\chi$ to the second order. Starting from this, Equation (\ref{eq5}) gives $\alpha_\chi$; Equations (\ref{eq1}) gives $\kappa_\chi$; Equation (\ref{eq2}) and (\ref{eq3}) give $\delta_\chi$ and $v_\chi$, respectively. Then, we have \begin{widetext}
\bea
   \alpha_\chi
   &=& - {3 \over 5} C
       + {3 \over 5} C^2
       + {9 \over 25} \Delta^{-1} \left[ - C \Delta C
       + 3 \Delta^{-1} \nabla^\alpha \nabla^\beta \left(
       C C_{,\alpha\beta} \right) \right]
       + {6 \over 175} {1 \over a^2 H^2}
       \left[ C^{,\alpha} C_{,\alpha}
       + 5 \Delta^{-1} \nabla^\alpha \left( C_{,\alpha} \Delta C \right) \right],
   \nonumber \\
   \varphi_\chi
   &=& {3 \over 5} C
       - {12 \over 25} C^2
       + {6 \over 25} \Delta^{-1} \left[ - C \Delta C
       + 3 \Delta^{-1} \nabla^\alpha \nabla^\beta \left(
       C C_{,\alpha\beta} \right) \right]
       - {6 \over 175} {1 \over a^2 H^2}
       \left[ C^{,\alpha} C_{,\alpha}
       + 5 \Delta^{-1} \nabla^\alpha \left( C_{,\alpha} \Delta C \right) \right],
   \nonumber \\
   v_\chi
   &=& {1 \over a H} \left\{
       - {2 \over 5} C
       + {7 \over 25} C^2
       + {6 \over 25} \Delta^{-1} \left[ - C \Delta C
       + 3 \Delta^{-1} \nabla^\alpha \nabla^\beta \left(
       C C_{,\alpha\beta} \right) \right] \right\}
       + {4 \over 175} {1 \over a^3 H^3}
       \left[ 2 C^{,\alpha} C_{,\alpha}
       + 3 \Delta^{-1} \nabla^\alpha \left( C_{,\alpha} \Delta C \right) \right],
   \nonumber \\
   \kappa_\chi
   &=& - {9 \over 5} H C
       + {9 \over 50} H C^2
       + {27 \over 25} H \Delta^{-1} \left[ - C \Delta C
       + 3 \Delta^{-1} \nabla^\alpha \nabla^\beta \left(
       C C_{,\alpha\beta} \right) \right]
       + {36 \over 175} {1 \over a^2 H}
       \left[ C^{,\alpha} C_{,\alpha}
       + 5 \Delta^{-1} \nabla^\alpha \left( C_{,\alpha} \Delta C \right) \right],
   \nonumber \\
   \delta_\chi
   &=& {6 \over 5} \left( 1 - {1\over 3} {\Delta \over a^2 H^2} \right) C
       + {6 \over 25} C^2
       + {18 \over 25} \Delta^{-1} \left[ C \Delta C
       - 3 \Delta^{-1} \nabla^\alpha \nabla^\beta \left(
       C C_{,\alpha\beta} \right) \right]
   \nonumber \\
   & &
       + {1 \over 175} {1 \over a^2 H^2} \left[
       224 C \Delta C + 81 C^{,\alpha} C_{,\alpha}
       - 36 \Delta^{-1} \nabla^\alpha \left( C_{,\alpha} \Delta C \right) \right]
       + {4 \over 175} {1 \over a^4 H^4} \left[
       \Delta \left( C^{,\alpha} C_{,\alpha} \right)
       + 5 \nabla^\alpha \left( C_{,\alpha} \Delta C \right) \right].
   \label{sol-ZSG}
\eea \end{widetext} These are a complete set of solutions in the zero-shear gauge. Note that $\alpha_\chi$ and $\varphi_\chi$ are presented in Boubekeur et al. (2009).

\subsection{Uniform-density gauge}

The uniform-density gauge takes \bea
   & & \delta \equiv 0 \equiv \gamma.
\eea
To the linear order, the solutions are \bea
   & & \varphi_\delta = \left( 1 - {2 \over 15} {\Delta \over a^2 H^2} \right) C,
       \quad
       v_\delta = - {2 \over 15} {1 \over a H}
       {\Delta \over a^2 H^2} C,
   \nonumber \\
   & &
       \kappa_\delta = - {1 \over H}
       \left( 1 - {2 \over 15} {\Delta \over a^2 H^2} \right) {\Delta \over a^2} C, \quad
       \alpha_\delta = - {1 \over 3} {\Delta \over a^2 H^2} C,
   \nonumber \\
   & &
       \chi_\delta = {2 \over 15} {1 \over H}
       \left( 3 - {\Delta \over a^2 H^2} \right) C.
\eea

To the second order, evaluating Equation (\ref{varphi_delta}) in the comoving gauge we can derive \begin{widetext} \bea
   \varphi_\delta
   &=& \left( 1 - {2 \over 15} {\Delta \over a^2 H^2} \right) C
       + {1 \over 5} {1 \over a^2 H^2} \left[
       {4 \over 3} C \Delta C
       + {1 \over 2} C^{,\alpha} C_{,\alpha}
       - \Delta^{-1} \nabla^\alpha \left( C_{,\alpha} \Delta C \right) \right]
   \nonumber \\
   & &
       + {2 \over 75} {1 \over a^4 H^4} \left\{
       {2 \over 7} \left[ \Delta \left( C^{,\alpha} C_{,\alpha} \right)
       + 5 \nabla^\alpha \left( C_{,\alpha} \Delta C \right) \right]
       + {1 \over 3} \left( \Delta C \right)^2
       + C^{,\alpha} \left( \Delta C_{,\alpha} \right)
       - \Delta^{-1} \nabla^\alpha \nabla^\beta
       \left[ C_{,\alpha} \left( \Delta C_{,\beta} \right) \right] \right\}
   \nonumber \\
   & &
       + {1 \over 225} {1 \over a^6 H^6} \left\{
       - \left( \Delta C^{,\alpha} \right) \Delta C_{,\alpha}
       + \Delta^{-1} \nabla^\alpha \nabla^\beta
       \left[ \left( \Delta C_{,\alpha} \right) \Delta C_{,\beta} \right] \right\}.
\eea \end{widetext} The rest of the solutions can be derived similarly: Equation (\ref{eq2}) gives $\kappa_\delta$; Equation (\ref{eq4}) gives $\alpha_\delta$; Equations (\ref{eq1}) and (\ref{eq6}) give $\chi_\delta$ and $v_\delta$, respectively.

\subsection{Uniform-expansion gauge}

The uniform-expansion gauge takes \bea
   & & \kappa \equiv 0 \equiv \gamma.
\eea
It is often called as the uniform-Hubble gauge.
To the linear order, the solutions are \bea
   & & \varphi_\kappa = {3 \dot H + {3 \over 5} {\Delta \over a^2}
       \over 3 \dot H + {\Delta \over a^2}} C, \quad
       \delta_\kappa = - {2 \over 3} {1 \over H^2}
       {3 \dot H + {3 \over 5} {\Delta \over a^2}
       \over 3 \dot H + {\Delta \over a^2}} {\Delta \over a^2} C,
   \nonumber \\
   & &
       \chi_\kappa = - {9 \over 5} {H \over 3 \dot H + {\Delta \over a^2}} C, \quad
       v_\kappa = - {2 \over 5} {1 \over a H}
       {1 \over 3 \dot H + {\Delta \over a^2}} {\Delta \over a^2} C,
   \nonumber \\
   & &
       \alpha_\kappa = - {3 \dot H + {3 \over 5} {\Delta \over a^2}
       \over \left( 3 \dot H + {\Delta \over a^2} \right)^2}
       {\Delta \over a^2} C.
\eea

To the second order, by evaluating Equation (\ref{varphi_kappa}) in the comoving gauge or the zero-shear gauge we can derive $\varphi_\kappa$. The rest of solutions can be derived as follows: Equation (\ref{eq2}) gives $\delta_\kappa$; Equation (\ref{eq4}) gives $\alpha_\kappa$; Equation (\ref{eq1}) gives $\chi_\kappa$; Equation (\ref{eq3}) gives $v_\kappa$. As the solutions are lengthy we omit presenting them. Instead, we present some asymptotic behaviors.
In the large-scale limit we have \bea
   & & \varphi_\kappa = C.
\eea
In the small-scale limit, we have \bea
   & & \varphi_\kappa = \varphi_\chi = -\alpha_\kappa = - \alpha_\chi, \quad
       \delta_\kappa = \delta_v,
   \nonumber \\
   & &
       {\Delta \over a} v_\kappa = {\Delta \over a} v_\chi
       = \kappa_v, \quad
       {\Delta \over a^2 H^2} \chi_\kappa = {9 \over 2} a v_\kappa.
   \label{solutions-UEG-SS-2}
\eea In the next section we will discuss these asymptotic solutions more.

\subsection{Uniform-curvature gauge}

The uniform-curvature gauge takes \bea
   & & \varphi \equiv 0 \equiv \gamma.
\eea It is often called as the flat gauge.
To the linear order, the solutions are \bea
   & & \delta_\varphi = \left( 3 - {2 \over 5} {\Delta \over a^2 H^2} \right) C, \quad
       \kappa_\varphi = 3 H \left( - {3 \over 2} + {1 \over 5}
       {\Delta \over a^2 H^2} \right) C,
   \nonumber \\
   & &
       \alpha_\varphi = - {3 \over 2} C, \quad
       \chi_\varphi = - {3 \over 5} {1 \over H} C, \quad
       v_\varphi = - {1 \over aH} C.
\eea
By evaluating Equation (\ref{delta_varphi}) in the comoving gauge or the zero-shear gauge we have \bea
   & & \delta_\varphi
       = \left( 3 - {2 \over 5} {\Delta \over a^2 H^2} \right) C
       + {3 \over 2} C^2
   \nonumber \\
   & & \qquad \!\!\!\!\!
       + {1 \over 5} {1 \over a^2 H^2} \left[
       4 C \Delta C
       + {9 \over 8} C^{,\alpha} C_{,\alpha}
       - {9 \over 4} \Delta^{-1} \nabla^\alpha \left( C_{,\alpha} \Delta C \right) \right]
   \nonumber \\
   & & \qquad \!\!\!\!\!
       + {4 \over 175} {1 \over a^4 H^4}
       \left[ \Delta \left( C^{,\alpha} C_{,\alpha} \right)
       + 5 \nabla^\alpha \left(
       C_{,\alpha} \Delta C \right) \right].
\eea The rest of the solutions can be derived as follows: Equation (\ref{eq2}) gives $\kappa_\varphi$; Equation (\ref{eq4}) gives $\alpha_\varphi$; Equation (\ref{eq1}) gives $\chi_\varphi$; Equation (\ref{eq3}) gives $v_\delta$.

\subsection{Large-scale evolution of curvature perturbation}
                                     \label{sec:conservations}

The curvature perturbations in various gauge conditions are known to be conserved in the large-scale limit. From our solutions, we have
\bea
   & & \varphi_v = \varphi_\delta = \varphi_\kappa = C,
   \nonumber \\
   & & \varphi_\chi = {3 \over 5} C
       - {12 \over 25} C^2
   \nonumber \\
   & & \qquad
       + {6 \over 25} \Delta^{-1} \left[ - C \Delta C
       + 3 \Delta^{-1} \nabla^\alpha \nabla^\beta \left(
       C C_{,\alpha\beta} \right) \right],
\eea valid to the second order. General conservation properties of $\varphi_v$, $\varphi_\delta$ and $\varphi_\kappa$ in the large-scale limit to nonlinear order are studied in Lyth et al. (2005) and Hwang \& Noh (2007). In addition, as we consider matter dominated era with fixed equation of state, $\varphi_\chi$ is also conserved. The following notations are often used in the literature \bea
   & & {\cal R} \equiv \varphi_v, \quad
       \zeta \equiv \varphi_\delta, \quad
       \Phi_A \equiv \alpha_\chi, \quad
       \Phi_H \equiv \varphi_\chi.
\eea $\Phi_A$ and $\Phi_H$ are often termed the Bardeen potential and Bardeen curvature, respectively (Bardeen 1980); these correspond to the Newtonian and the post-Newtonian gravitational potentials, respectively (Hwang et al. 2008).

\section{Relativistic/Newtonian correspondences}
                                        \label{sec:correspondences}

In this section we restore the speed of light $c$; in our convention the scale factor has a length dimension. In Newtonian context, the mass conservation, momentum conservation, and the Poisson's equation give (Peebles 1980) \bea
   & & \dot \delta
       + {1 \over a} \nabla \cdot {\bf u}
       = - {1 \over a} \nabla \cdot \left( \delta {\bf u} \right),
   \label{dot-delta-eq-N} \\
   & & \dot {\bf u} + H {\bf u}
       + {1 \over a} \nabla \Phi
       = - {1 \over a} {\bf u} \cdot \nabla {\bf u},
   \label{dot-u-eq-N} \\
   & & {\Delta \over a^2} \Phi
       = 4 \pi G \varrho \delta,
   \label{Poisson-eq-N}
\eea where $\delta$, ${\bf u}$, and $\Phi$ are the relative mass density perturbation $\delta \varrho / \varrho$, the velocity perturbation, and the perturbed gravitational potential, respectively. By removing the gravitational potential Equations (\ref{dot-u-eq-N}) and (\ref{Poisson-eq-N}) give \bea
   & & {1 \over a} \nabla \cdot \left( \dot {\bf u} + H {\bf u} \right)
       + 4 \pi G \varrho \delta
       = - {1 \over a^2} \nabla \cdot \left( {\bf u} \cdot \nabla {\bf u} \right).
   \label{dot-u-eq-N-2}
\eea In Newtonian context these equations are valid to fully nonlinear order.

In Einstein's gravity, in the comoving gauge, from Equations (\ref{eq6}) and (\ref{eq7}), Equations (\ref{eq4}) and (\ref{eq7}), and Equation (\ref{eq3}), respectively, we have \bea
   & & \dot \delta_v - \kappa_v
       = \kappa_v \delta_v - {c \over a^2} \delta_{v,\alpha} \chi_v^{\;,\alpha},
   \label{RN1} \\
   & & \dot \kappa_v + 2 H \kappa_v
       - 4 \pi G \varrho \delta_v
       = {1 \over 3} \kappa_v^2
       - {c \over a^2} \kappa_{v,\alpha} \chi_v^{\;,\alpha}
   \nonumber \\
   & & \qquad
       + {c^2 \over a^4} \left[ \chi_v^{\;,\alpha\beta} \chi_{v,\alpha\beta}
       - {1 \over 3} \left( \Delta \chi_v \right)^2 \right],
   \label{RN2} \\
   & & \kappa_v = - {\Delta \over a^2} \chi_v
       + n_2 |_v.
   \label{RN3}
\eea By the identifications \bea
   & & \delta_v = \delta, \quad
       \kappa_v \equiv - {1 \over a} \nabla \cdot {\bf u},
   \label{RN-correspondence}
\eea to the second order, and \bea
   & & {c \over a} \nabla \chi_v = - \nabla v_\chi \equiv {\bf u}
       \equiv \nabla u,
   \label{RN-correspondence-2}
\eea to the linear order, Equations (\ref{RN1}) and (\ref{RN2}) {\it exactly coincide} with Equations (\ref{dot-delta-eq-N}) and (\ref{dot-u-eq-N-2}), respectively. We have termed this coincidence, a relativistic/Newtonian correspondence of zero-pressure fluid in flat background to the second order in perturbation (NH2004, Noh \& Hwang 2005). Therefore, solutions for $\delta_v$ and $\kappa_v$ are valid for the Newtonian $\delta$ and ${\bf u}$, respectively, to the second order.

From Equation (\ref{Poisson-eq-N}) with $\delta = \delta_v$ to the second order, we can construct relativistic combination of gravitational potential which exactly reproduces the Newtonian gravitational potential to the second order. From Eqs.\ (\ref{eq2}), (\ref{eq3}) and (\ref{eq5}), and using Equation (\ref{delta_v}), we have \begin{widetext}
\bea
   & & {4 \pi G \varrho \over c^2} \delta_v
       + {\Delta \over a^2} \varphi_\chi
       = {\Delta \over a^2} \left(
       2 \alpha_\chi^2
       - {2 \pi G \varrho \over c^4} a^2 v_\chi^2 \right)
       - {5 \over 2 a^2} \alpha_\chi^{\;,\alpha} \alpha_{\chi,\alpha}
       + {12 \pi G \varrho \over c^4} a H \Delta^{-1} \nabla^\alpha
       \left( v_\chi \delta_{v,\alpha} \right),
   \\
   & & \alpha_\chi + \varphi_\chi
       = 2 \alpha_\chi^2
       + \Delta^{-1} \left( \alpha_\chi^{\;,\alpha} \alpha_{\chi,\alpha}
       + {4 \pi G \varrho \over c^4} a^2 v_\chi^{\;,\alpha} v_{\chi,\alpha} \right)
       - 3 \Delta^{-2} \nabla^\alpha \nabla^\beta
       \left( \alpha_{\chi,\alpha} \alpha_{\chi,\beta}
       + {4 \pi G \varrho \over c^4} a^2 v_{\chi,\alpha} v_{\chi,\beta} \right),
\eea
where we have used Equations (\ref{eq3}), (\ref{eq6}), (\ref{eq7}) to the linear order. We have assumed only $K = 0$. Using these we can express the relativistic combination of the gravitational potential to the second order using Newtonian variables as \bea
   & & \Phi_{\rm GR}
       = 4 \pi G \varrho a^2 \Delta^{-1} \delta_v
       = - c^2 \varphi_\chi
       + {1 \over c^2} \Big[ 2 \Phi^2
       - 2 \pi G \varrho a^2 u^2
       - 12 \pi G \varrho a^3 H \Delta^{-2} \nabla \cdot \left( u \nabla \delta \right) - {5 \over 2} \Delta^{-1} \left( \nabla \Phi \right)^2 \Big]
   \nonumber \\
   & & \qquad\;
       = c^2 \alpha_\chi
       + {1 \over c^2} \Big\{ - 2 \pi G \varrho a^2 u^2
       - 12 \pi G \varrho a^3 H \Delta^{-2} \nabla \cdot \left( u \nabla \delta \right)
       - \Delta^{-1} \left[ {7 \over 2} \left( \nabla \Phi \right)^2
       + 4 \pi G \varrho a^2 {\bf u}^2 \right]
   \nonumber \\
   & & \qquad \quad
       + 3 \Delta^{-2} \nabla^\alpha \nabla^\beta \left(
       \Phi_{,\alpha} \Phi_{,\beta}
       + 4 \pi G \varrho a^2 u_\alpha u_\beta
       \right)
       \Big\}.
   \label{Phi_GR}
\eea \end{widetext}
Examination of the quadratic-order correction terms shows that those can be regarded as the post-Newtonian order corrections with ${1 \over c^2} \Phi \sim {aH \over c^2} u \sim {a^2 H^2 \over c^2 \Delta} \delta$ order smaller than the leading order term, and become negligible in the small-scale limit. Therefore, with identifications of $\delta_v$, $\kappa_v$ and $\Phi_{\rm GR}$ given in Equations (\ref{RN-correspondence}), (\ref{RN-correspondence-2}) and (\ref{Phi_GR}) as the Newtonian density, velocity and gravitational potential fluctuations, respectively \bea
   & & \delta_v = \delta, \quad
       \kappa_v \equiv - {1 \over a} \nabla \cdot {\bf u}, \quad
       \Phi_{\rm GR} = \Phi,
   \label{RN-correspondence-exact}
\eea we have exact relativistic/Newtonian correspondence to the second order with Equations (\ref{dot-delta-eq-N})-(\ref{Poisson-eq-N}).

For Newtonian gravitational potential, to the second order, we have \bea
   & & {1 \over c^2} \Phi
       = {3 \over 2} {a^2 H^2 \over c^2 \Delta} \delta
       = - {3 \over 5} C
       + {9 \over 20} C^2
       + {3 \over 2} \Delta^{-1} \left( C \Delta C \right)
   \nonumber \\
   & & \qquad
       + {6 \over 175} {c^2 \over a^2 H^2}
       \left[ C^{,\alpha} C_{,\alpha}
       + 5 \Delta^{-1} \nabla^\alpha \left( C_{,\alpha} \Delta C \right) \right].
\eea This, in general, differs from $\alpha_\chi$ which also differs from $- \varphi_\chi$ as can be seen in Equation (\ref{Phi_GR}). However, to the linear order in general, or to the second order in the small-scale limit, we have \bea
   & & \alpha_\chi = - \varphi_\chi = {1 \over c^2} \Phi,
\eea where $\alpha_\chi$ and $\varphi_\chi$ correspond to the Newtonian and the post-Newtonian gravitational potentials, respectively (Hwang et al. 2008). Therefore, in the sub-horizon limit, even to the second order we have \bea
   & & \delta_v, \quad
       \kappa_v ( = \Delta v_\chi / a ), \quad
       \alpha_\chi (= - \varphi_\chi),
\eea which correspond to the Newtonian perturbations $\delta$, $- {1 \over a} \nabla \cdot {\bf u}$ and $\Phi$, respectively. In the following we present further correspondences available in other gauge conditions, especially in the small-scale limit.

In the small-scale limit, from our complete solutions we notice that \begin{widetext} \bea
   & & \delta_v
       = \delta_\chi
       = \delta_\kappa
       = \delta_\varphi
       = - {2 \over 5} {c^2 \Delta \over a^2 H^2} C
       + {4 \over 175} {c^4 \over a^4 H^4}
       \left[ \Delta \left( C^{,\alpha} C_{,\alpha} \right)
       + 5 \nabla^\alpha \left(
       C_{,\alpha} \Delta C \right) \right],
   \nonumber \\
   & &
   \kappa_v
       = - c {\Delta \over a^2} \chi_v
       = {\Delta \over a} v_\chi
       = {\Delta \over a} v_\kappa
       = - {2 \over 5} {c^2 \Delta \over a^2 H} C
       + {4 \over 175} {c^4 \over a^4 H^3}
       \left[ 2 \Delta \left( C^{,\alpha} C_{,\alpha} \right)
       + 3 \nabla^\alpha \left(
       C_{,\alpha} \Delta C \right) \right],
   \nonumber \\
   & & \alpha_\chi
       = - \varphi_\chi
       = \alpha_\kappa
       = - \varphi_\kappa
       = - {3 \over 5} C
       + {6 \over 175} {c^2 \over a^2 H^2}
       \left[ C^{,\alpha} C_{,\alpha}
       + 5 \Delta^{-1} \nabla^\alpha \left( C_{,\alpha} \Delta C \right) \right].
   \label{SS-identification}
\eea \end{widetext} Thus, we have \bea
   & & \delta_v = {2 \over 3} {c^2 \Delta \over a^2 H^2} \alpha_\chi,
   \label{Poisson-eq-E}
\eea which is the relativistic version of Poisson's equation; it follows from Equations (\ref{eq2}), (\ref{eq3}) and (\ref{eq5}) for $K = 0$. In Equation (\ref{Poisson-eq-E}) instead of $\delta_v$ and $\alpha_\chi$ we can replace other variable identified in Equation (\ref{SS-identification}) as well. We have shown that $\delta_v$ and $\kappa_v$ exactly correspond to the Newtonian density and velocity perturbations, whereas $\alpha_\chi = - \varphi_\chi$ corresponds to Newtonian potential perturbation in the small-scale limit.

To the linear order, the perturbation variables $\delta_v$, $v_\chi$ and $\alpha_\chi (= - \varphi_\chi)$ are known to correspond to the Newtonian relative density perturbation, velocity perturbation and the gravitational potential perturbation, respectively (Bardeen 1980). Also to the linear order, all perturbation variables of density, velocity, and potential in the zero-shear gauge and the uniform-expansion gauge are known to have Newtonian correspondences in the sub-horizon limit: see Section 84 in Peebles (1980), and Table I in Hwang \& Noh (1999).
In this work we show that the same correspondences are now valid to the second order in the sub-horizon scale. We have the Newtonian correspondences for the following variable in the sub-horizon scale \bea
   & &
       \delta
       \equiv \delta_\kappa
       = \delta_\chi
       = \delta_\varphi
       = \delta_v,
   \nonumber \\
   & &
       - {1 \over a} \nabla \cdot {\bf u}
       \equiv {\Delta \over a} v_\kappa
       = {\Delta \over a} v_\chi
       = \kappa_v,
   \nonumber \\
   & &
       {1 \over c^2} \Phi
       \equiv \alpha_\kappa
       = \alpha_\chi
       = - \varphi_\kappa
       = - \varphi_\chi.
   \label{solutions-UEG-SS}
\eea
We can summarize the relativistic/Newtonian correspondences in a table form as the following: \begin{widetext}
\begin{center}
\begin{tabular}{l|l|l|l|l}
    \hline \hline
        \hskip 3cm & CG \qquad\quad & ZSG & UEG & UCG \quad\quad \\
    \hline \hline
        Exact: Second order & $\delta_v$, $\kappa_v$ &  &  &  \\
    \hline
        \hskip 1.17cm Linear order & $\delta_v$, $\kappa_v$ & \hskip .57cm $v_\chi$, $\alpha_\chi (= - \varphi_\chi)$ &  &  \\
    \hline \hline
        Small-scale: Second order & $\delta_v$, $\kappa_v$ & $\delta_\chi$, $v_\chi$, $\alpha_\chi (= - \varphi_\chi)$ & $\delta_\kappa$, $v_\kappa$, $\alpha_\kappa (= - \varphi_\kappa)$ & $\delta_\varphi$ \\
    \hline
        \hskip 1.95cm Linear order & $\delta_v$, $\kappa_v$ & $\delta_\chi$, $v_\chi$, $\alpha_\chi (= - \varphi_\chi)$ & $\delta_\kappa$, $v_\kappa$, $\alpha_\kappa (= - \varphi_\kappa)$ & $\delta_\varphi$ \\
    \hline \hline
\end{tabular}
\end{center}
\hskip 2cm{Table 1. Relativistic/Newtonian correspondences available in different gauge conditions.}
\end{widetext}
The CG, ZSG, UEG, and UCG are acronyms of the comoving gauge, the zero-shear gauge, the uniform-expansion gauge, and the uniform-curvature gauge, respectively. Although the complete small-scale correspondences to the second order in the zero-shear gauge and the uniform-expansion gauge are impressive, lack of such correspondences in the other gauge indicates that the correspondence is still a non-trivial consequence available only under certain gauge conditions.

We note that in our spatial gauge condition ($\gamma \equiv 0$), the (temporal) comoving gauge ($v \equiv 0$) does not imply the (temporal) synchronous gauge ($\alpha \equiv 0$). The subtle relations between our comoving gauge ($v \equiv 0 \equiv \gamma$) and the original synchronous gauge ($\alpha \equiv 0 \equiv \beta$) to the second order are studied in Hwang \& Noh (2006).

A complementary study of the correspondence between the linear perturbation theory and the post-Newtonian approximation is made in Noh \& Hwang (2012).

\section{Discussion}
                                     \label{sec:discussion}

In this work we have presented the gauge transformation properties and several widely used gauge-invariant combinations to the second order in the context of a general fluid: see Section \ref{sec:GT}. In the main part we have presented the growing mode solutions of a zero-pressure fluid to the second order in several fundamental gauge conditions: see Section \ref{sec:solutions}. Based on the solutions we have clarified the relativistic/Newtonian correspondences available in different gauge conditions: see Section \ref{sec:correspondences}. We have newly identified the gauge-invariant combinations of relativistic perturbation variables which exactly reproduce the perturbed Newtonian gravitational potential to the second order: see Equation (\ref{Phi_GR}). Together with the previously identified gauge-invariant density and velocity perturbations variables, now we complete the exact relativiatic/Newtonian correspondence to the second order perturbation: see Equation (\ref{RN-correspondence-exact}).

In Chisari \& Zaldarriaga (2011) and Green \& Wald (2011) the authors have considered situation where large-scale (near horizon-scale) fluctuations are linear whereas local density inhomogeneity is nonlinear. Both analyses were made in the zero-shear gauge (often termed the longitudinal gauge or the conformal Newtonian gauge). These two works are complementary to our present work in addressing relativistic identifications of Newtonian perturbations (the relativistic/Newtonian correspondences): their works cover large-scale linear perturbation theory in the presence of small-scale nonlinear density inhomogeneity, whereas our work treats the second-order perturbations in all scales self-consistently. In the overlapping regime the results coincide. Both works have identified velocity and gravitational potential perturbations in the zero-shear gauge ($v_\chi$ and $\alpha_\chi$) and density perturbations in the comoving gauge ($\delta_v$) as the corresponding Newtonian perturbation variables: this coincides with our identification to the linear order. For our extension of the identifications exactly valid to the second order, see Equation (\ref{Phi_GR}).

Extensions of the solutions in this work to cases with (i) general $K$ and $\Lambda$, (ii) decaying solutions, (iii) general pressure, (iv) vector and tensor perturbations, (v) third-order perturbations, etc., are left for future studies. For our purpose of clarifying the relativistic/Newtonian correspondence our present study is enough (except for assuming $\Lambda$) as the correspondence in Equation (\ref{Phi_GR}) is valid up to the second order only for $K = 0$. Although the growing mode solutions in Section \ref{sec:solutions} are presented for $\Lambda = 0$, the correspondences in Equation (\ref{Phi_GR}) are valid in the equation level with general $\Lambda$.

%
%
\acknowledgments
J.G.\ was supported in part by a Korean-CERN fellowship.
H.N.\ was supported by grants No.\ C00022 from the Korea Research Foundation (KRF) and No.\ 2009-0078118 from KOSEF funded by the Korean Government (MEST).
J.H.\ was supported by KRF Grant funded by the Korean Government
(KRF-2008-341-C00022).

%
%

\vspace{1cm}\noindent CERN-PH-TH/2012-004

\end{document}